\documentclass[11pt,showpacs,groupedaddress]{revtex4}
\makeatletter
\usepackage{calc}
\usepackage{times}
\usepackage{amsmath, amssymb}
\usepackage{multirow}
\def\be{\begin{equation}}
\def\ee{\end{equation}}
\def\bea{\begin{eqnarray}}
\def\eea{\end{eqnarray}}
\def\bma{\begin{mathletters}}
\def\ema{\end{mathletters}}
\def\bi{\begin{itemize}}
\def\ei{\end{itemize}}

\def\C{\hbox{$\mit I$\kern-.7em$\mit C$}}

\newcommand{\singlespacing}{\let\CS=\@currsize\renewcommand{\baselinestretch}
{1.0}\tiny\CS}
\newcommand{\doublespacing}{\let\CS=\@currsize\renewcommand{\baselinestretch}
{1.5}\tiny\CS}

\begin{document}

\title{A complementary relation between classical bits and randomness in local part \\in simulating singlet state}

\author{Guruprasad Kar}
\author{MD. Rajjak Gazi}
\author{Manik Banik}
\affiliation{Physics and Applied Mathematics Unit, Indian
Statistical Institute, 203 B.T. Road, Kolkata-700108, India}

\author{Subhadipa Das}
\author{Ashutosh Rai}
\affiliation{S.N.Bose National Center for Basic Sciences, Block
JD, Sector III, Salt Lake, Kolkata-700098, India}

\author{Samir Kunkri}
\affiliation{Mahadevananda Mahavidyalaya, Monirampore,
Barrackpore, North 24 Parganas, 700120, India }

\begin{abstract}
Recently simulating the statistics of singlet state with non-quantum resources
has generated much interest. Singlet
state statistics can be simulated by $1$ bit of classical
communication without using any further nonlocal correlation. But,
interestingly, singlet state statistics can also be simulated with
no classical cost if a non-local box is used. In the first case,
the output is completely biased whereas in second case outputs are
completely random. We suggest a new (possibly) signaling
correlation resource which successfully simulates singlet
statistics and this result suggests a complementary relation
between required classical bits and randomness in local output
involved in the simulation. Our
result reproduces the  above two models of simulation as extreme
cases. This also suggests another important feature in Leggett's non-local model and the
model presented by Branciard \emph{et.al.}
\end{abstract}
\pacs{03.65.Nk, 03.65.Yz}

\maketitle
\section{Introduction}
Violation of Bell's inequality \cite{bell} by quantum statistics generated from
singlet state implies impossibility of reproducing all quantum results by local hidden variable
theory. Then Leggett proposed a non-local hidden variable model
with some constraint on local statistics and showed that this
model is incompatible with quantum mechanics
\cite{leggett,groblacher}. The result was further generalized by
Branciard \emph{et.al.} \cite{branciard}. All these results have
generated a new interest in simulating singlet statistics by some
non-local correlation. In this context, it should be mentioned
that if one cbit of communication is allowed, the singlet
statistics can be simulated \cite{toner}. After this work, quite
interestingly, singlet statistics was simulated without
communication by using the Popescu-Rorlich (PR) Box \cite{cerf}.
Recently Colbeck and Renner \cite{colbeck} proved a general result
by showing that no non-signalling non-local model can generate
statistics of singlet state if the the model has non-trivial local part
and this result is deeply related to the simulation problem.
This result was further supported by the work of Branciard \emph{et.al.} \cite{branciard}.

 Here, in this work, we suggest a general (possibly) signaling
 correlation which can be seen as convex combination of a correlation with communication capacity of 1 bit
 and a PR box. We show that with this type of signalling
 correlation singlet statistics can be generated. Our result suggests  a complementary relation between the amount of classical
 communication required and randomness in the local binary output in the task of simulating singlet
 correlation with classical communication which is limited by $1$ cbit.
 \section{Correlation embracing classical communication}
 In order to produce our result we consider the following correlation (hereafter designated by  $S^p$),
 with binary input $x,y\in \{0,1\}$ and binary output $a,b\in
 \{0,1\}$,
\begin{equation}
 P(ab|xy)= (xy\oplus\delta_{ab})[(a\oplus1)p+a(1-p)]
 \end{equation}
where $P(ab|xy)$ is probability of getting
outputs $a$ and $b$ for corresponding inputs $x$ and $y$, and $\frac{1}{2} \leq p \leq 1$.
 Here and from now on, $\oplus$ represents addition modulo $2$ and $\delta_{ab}=a\oplus b\oplus 1$.
 Interestingly for $p \neq \frac{1}{2}$, $S^p$ correlation violates no-signaling. In particular,
 for $p = 1$, this correlation designated by  $S^{1 cbit}$ can be used to communicate 1 cbit from Alice to Bob.
 For $p = \frac{1}{2}$, this is a PR box written as $P^{NL}$.
 Then it can be easily shown that
\begin{equation}
 S^p = (2p - 1)S^{1 cbit} + 2(1 - p)P^{NL}.
 \end{equation}
 \section{Simulation of singlet by correlation $\mathbf{S^p}$}
The protocol for simulating the singlet state statistics by correlation $S^p$ is same as given in \cite{cerf}.
For completeness we briefly describe the protocol. Alice and Bob share the correlation $S^p$ along with  shared
randomness in the forms of pairs of normalized vectors
 $\hat{\lambda}_1$ and  $\hat{\lambda}_2$, randomly and independently distributed  over the Poincare sphere.
 When simulating singlet, let in one turn Alice and Bob  have been asked to provide the result of
 measurements along (unit) vectors $\hat{A}$ and $\hat{B}$ respectively.
 The protocol runs as follows. Alice calculates the following quantity

\begin{equation}
x = \mbox{sgn}(\hat{A}\cdot\hat{\lambda}_1 )\oplus
\mbox{sgn}(\hat{A}\cdot\hat{\lambda}_2)
\end{equation}
and inserts it as input to the machine ($S^p$ correlation), where,
\[ \mbox{sgn}(z) = \left\{ \begin{array}{ll}
         1 & \mbox{if}~~ z\geq 0;\\
        0 & \mbox{if}~~ z<0.\end{array} \right. \]

As result of  measurement, Alice provide the following quantity
\begin{equation}
v(\hat{A}) = a\oplus \mbox{sgn}(\hat{A}\cdot\hat{\lambda}_1)
 \end{equation}
 $a$ being the output from the machine.
Bob calculates the following quantity
\begin{equation}
y = \mbox{sgn}(\hat{B}\cdot\vec{\lambda}_+) \oplus
\mbox{sgn}(\hat{B}\cdot\vec{\lambda}_-),
 \end{equation}
where $\vec{\lambda}_{\pm}=\hat{\lambda}_1\pm
\hat{\lambda}_2$ and insert this as input to the machine. After receiving the bit $b$ from the
machine, he provide the following as measurement result
\begin{equation}
v(\hat{B}) = b\oplus
\mbox{sgn}(\hat{B}\cdot\vec{\lambda}_+)\oplus1 .
 \end{equation}
Armed with the correlation $S^p$ one can easily apply the same strategy for simulating singlet correlation
 \begin{equation}
 E[ v(\hat{A})\oplus v(\hat{B}) | \hat{A}, \hat{B} ] = \frac{1 + \hat{A} \cdot \hat{B}}{2},
\end{equation}
 in the same line as in \cite{cerf}. To see how it works one should observe that, from
\begin{equation}
v(\hat{A})\oplus v(\hat{B}) = a \oplus b \oplus
\mbox{sgn}(\hat{A}\cdot\hat{\lambda}_1) \oplus
\mbox{sgn}(\hat{B}\cdot\vec{\lambda}_+)\oplus1 ,
 \end{equation}
using the correlation $S^p$ we get
 \begin{eqnarray}
  v(\hat{A})\oplus v(\hat{B})= [(2p-1)xy+2(1-p)xy] \oplus \mbox{sgn}(\hat{A}\cdot\hat{\lambda}_1)
  \oplus& \mbox{sgn}(\hat{B}\cdot\vec{\lambda}_+)\oplus1 \nonumber\\= xy \oplus \mbox{sgn}(\hat{A}\cdot\hat{\lambda}_1) \oplus
\mbox{sgn}(\hat{B}\cdot\vec{\lambda}_+)\oplus1
  \end{eqnarray}
 which is identical to the equation (10) in \cite{cerf} and the result immediately follows.
\section{A complementary relation}
 The $S^p$ correlation used in our model for simulating singlet introduces a biasness in the local output $R(p)$
 which is quantified by Shannon entropy of the outputs for a given input,
 \begin{equation}
 R(p) = H(p) = - p\log{p} - (1-p)\log{(1-p)}.
\end{equation}
 On the other hand, the amount of bits $C(p)$ that can be communicated from Alice to Bob by using $S^p$ correlation is quantified by
 the maximal mutual information between Alice's input and Bob's output (for Bob's input $1$) and this can be expressed as
\begin{equation}
C(p) = \max_{Alice's~~ input} I(x:b) = 1-H(p),
\end{equation}
where $ I(x:b)=H(x) + H(b|y=1)-H(xb|y=1)$. For Bob's input 0, the corresponding mutual information vanishes.
Hence we see that in simulating singlet statistics, as
communication capacity of the correlation resource increases, the
randomness of local output decreases and vice-versa. The
complementary relation for this model of simulation where the
classical communication is limited by $1$ cbit can be expressed by
\begin{eqnarray}
\mbox{Randomness in local output} + \mbox{Communication capacity
of the} \nonumber \\ \mbox{resource in use}~= R(p) + C(p) = 1.
\end{eqnarray}
Obviously one extreme point ($p = 1$) generates the Toner-Bacon model \cite{toner} and the other extreme point ($p =
 \frac{1}{2}$) generates the model presented by Cerf \emph{et.al.} \cite{cerf}.
 One should also note that the protocol for the simulation of the singlet by $S^{p}$ correlation does not depend on the value of $p$.
 Hence for the simulation, in general one can also use randomly chosen $S^{p}$ boxes from an ensemble $\{S^{p}: \frac{1}{2}\leq p \leq 1 \}$  where boxes with arbitrary $p$ labels appear according to some probability distribution $\rho (p)$. In this general picture, the complementary relation of the form $(13)$ still holds. Here
 average randomness and average communication are given by $\overline{R}=\int H(p) \rho (p) dp$ and $\overline{C}=\int C(p) \rho (p) dp$ respectively. Then by using the relation $(13)$ we get,
\begin{equation}
\overline{R}+\overline{C}=\int [H(p)+ C(p)] \rho (p) dp = 1.
\end{equation}

One must observe that in this model of simulation, the biasness of the measurement results for any given observable depends only on the
parameter of the non-local resources.
If this is not the case i.e. if in some model, biasness is a function of the direction of the observable,
then $R(p)$ can be taken as average of Shannon entropy of measurement results
over all possible measurements. Then the local randomness is given by
$R = \langle H(a_i)\rangle$, $H(a_i)$ being the Shannon entropy for outcome
for measurement along the direction $a_i$ on either Alice's side or Bob's side
(in all the models till considered, $R$ has been taken to be same on both side).

Now in the context of following  results: (i) simulating singlet state without communication requires complete randomness for local outcomes \cite{colbeck, branciard}, (ii) the absence of any less than 1 cbit protocol for simulating singlet state non-asymptotically using classical communications as the only nonlocal resource, and (iii) the complementary relation obtained in this work;
we conjecture that if there is a model for simulating statistics of singlet state
with the help of classical communication of $C$ bit on average, then the complementary relation $ R + C \geq 1$ holds as a necessary condition.
\section{Implications for Leggett's model}
 Next we apply our result to Leggett's model
\cite{leggett,groblacher}. In Leggett's non-local hidden variable
model, the local statistics for a given value of hidden variable
has been considered to be same as generated by some completely
polarized state and it has been shown that this model does not
reproduce singlet statistics. This result has been generalized in
\cite{branciard} where local statistics could also be generated by some
mixed polarized state and this also does not work for singlet simulation. In both these models, the local randomness is not
uniform and $R$ has to be calculated by taking average of Shannon
entropy of outcomes over all possible measurements performed on a
pure polarized state or a mixed polarized state on either side.
For a general mixed state $\rho = \frac{1}{2}[I + \mu
\vec{n}\cdot\vec{\sigma} ]$ with $0 \leq \mu \leq 1$, the
average entropy of output $R$ over all possible polarization
measurement is obtained as
\begin{eqnarray}
R = \langle H(a_i) \rangle =1-\frac{\left[(1
+\mu)^2\ln{(1+\mu)}-(1-\mu)^2\ln{(1-\mu)}
-2\mu \right]}{4\mu \ln{2}}.
\end{eqnarray}
From the above expression one can easily check that for $\mu \neq
0$, $R<1$ and the complementary relation (13) tells that the Leggett's
model \cite{leggett,groblacher} should fail to reproduce the statistics of singlet state, as no classical communication is used (Leggett's model is non-signaling).

Still one may question why there is a successful (non-signaling)
non-local model which reproduce singlet statistics for restricted
choice of observable \cite{supple}. This is possible because, for a given pure polarized
state, one can always choose the measurements in a plane of the
Poincare sphere which is orthogonal to the direction of
polarization and in that case $R=1$. Using our model of simulation of singlet, we can extend this result for the choice of observable $\hat{a}$ and $\hat{b}$ restricted on two cones $\hat{u}\cdot \hat{a}= \cos \theta$ and $\hat{v}\cdot \hat{b}=\cos \theta$ respectively where $\hat{u}$ and $\hat{v}$ represent the  directions of polarization of local states in the Leggett's model. But this would require ($1 - H(\cos ^2(\frac{\theta}{2}))$) bits of classical communication.
With the average local randomness $R\neq 1$ and further satisfying Malu's law \cite{leggett,groblacher} for arbitrary choices of observable for given polarized states on both sides,
whether singlet statistics can be simulated with the assistance of $1-R$ classical bit or even with finite amount of bits remains open.
We think that this is a qualitatively severe constraint (Malu's law) on local statistics and even communication of finite amount of bits may not work.

\emph{Note added}- After we finish this work we saw a  similar conjecture
proposed by Michael J.W. Hall \cite{hall}.

\acknowledgments
SD and AR acknowledge support from the DST project SR/S2/PU-16/2007.


\section*{References}
\begin{thebibliography}{99}

\bibitem{bell} J.S. Bell, Physics {\bf 1}, 195 (1964).
\bibitem{leggett} A. J. Leggett, Found. Phys. {\bf 33}, 1469 (2003).
\bibitem{groblacher} S. Groblacher, T. Paterek, R. Kaltenbaek,
C.Brukner, M. Zukowski, M. Aspelmeyer, and A. Zeilinger, Nature
{\bf 446}, 871 (2007).
\bibitem{branciard} C. Branciard, N. Brunner, N. Gisin, C. Kurtsiefer, A. Lamas-Linares, A. Ling, and V. Scarani , Nature physics
{\bf 4}, 681 (2008).
\bibitem{toner} B.F. Toner and D. Bacon, Phys. Rev. Lett. {\bf 91}, 187904 (2003).
\bibitem{cerf} N.J. Cerf, N. Gisin, S. Massar and S. Popescu, Phys. Rev. Lett. {\bf 94}, 220403 (2005).
\bibitem{colbeck} R. Colbeck and R. Renner, Phys. Rev. Lett. {\bf 101}, 050403 (2008).
\bibitem{supple} S. Groblacher, T. Paterek, R. Kaltenbaek,
C.Brukner, M. Zukowski, M. Aspelmeyer, and A. Zeilinger, Nature
{\bf 446}, 871 (2007) [Supplementary Information].
\bibitem{hall} Michael J.W. Hall, Phys. Rev. A {\bf 82}, 062117 (2010); Also at arXiv:1006.3680.
\end{thebibliography}
\end{document}